\documentclass[aps,prl,superscriptaddress,showpacs,amsmath, amssymb,floatfix,twocolumn]{revtex4-1}
\usepackage{graphicx, nicefrac}
\usepackage{color, colortbl, hyperref}
	\definecolor{LinkColor}{rgb}{0.45,0,0}
	\definecolor{UrlColor}{rgb}{0,0,0.45}
	\definecolor{CiteColor}{rgb}{0,0.45,0}
	\hypersetup{%
	linkcolor=LinkColor,%
	filecolor=LinkColor,%
	menucolor=LinkColor,%
	citecolor=CiteColor,%
	urlcolor=UrlColor,%
	colorlinks=true%
	}

\begin{document}
\title{Magnetic skyrmions and skyrmion clusters in the helical phase of Cu$_2$OSeO$_3$}

\author{Jan M\"uller}
\affiliation{Institute for Theoretical Physics, University of Cologne, D-50937 Cologne, Germany}
\author{Jayaraman Rajeswari}
\affiliation{Laboratory for Ultrafast Microscopy and Electron Scattering (LUMES), Institute of Physics, EPFL, CH-1015 Lausanne, Switzerland}
\author{Ping Huang}
\affiliation{Laboratory for Quantum Magnetism (LQM), Institute of Physics, EPFL, CH-1015 Lausanne, Switzerland}
\author{Yoshie Murooka}
\affiliation{Laboratory for Ultrafast Microscopy and Electron Scattering (LUMES), Institute of Physics, EPFL, CH-1015 Lausanne, Switzerland}
\author{Henrik M. R\o nnow}
\affiliation{Laboratory for Quantum Magnetism (LQM), Institute of Physics, EPFL, CH-1015 Lausanne, Switzerland}
\author{Fabrizio Carbone}
\affiliation{Laboratory for Ultrafast Microscopy and Electron Scattering (LUMES), Institute of Physics, EPFL, CH-1015 Lausanne, Switzerland}
\author{Achim Rosch}
\email{rosch@thp.uni-koeln.de}
\affiliation{Institute for Theoretical Physics, University of Cologne, D-50937 Cologne, Germany}

\date{\today}

\begin{abstract}
Skyrmions are nanometric spin whirls that can be stabilized in magnets lacking inversion symmetry. 
The properties of isolated skyrmions embedded in a ferromagnetic background have been intensively studied. We show that single skyrmions and clusters of skyrmions can also form in the helical phase and investigate theoretically their energetics and dynamics. 
The helical background provides natural one-dimensional channels along which a skyrmion can move rapidly. 
In contrast to skyrmions in ferromagnets, the skymion-skyrmion interaction has a strong attractive component and thus skyrmions tend to form clusters with characteristic shapes. 
These clusters are directly observed in transmission electron microscopy measurements in thin films of Cu$_2$OSeO$_3$. 
Topological quantization, high mobility and the confinement of skyrmions in channels provided by the helical background may be useful for future spintronics devices.
\end{abstract}


\maketitle

In crystals lacking inversion symmetry, subtle relativistic interactions lead to the formation of exotic spin textures \cite{dzyaloshinskii1964theory,bogdanov1989,bogdanov1994,janson2014quantumnature}. For example, in chiral magnets such as Cu$_2$OSeO$_3$, ferrimagnetically ordered spins spontaneously cant to form helices with a pitch of $60\--70$\;nm \cite{adams2012long,seki2012formation,rajeswari2015filming}. Upon applying a tiny magnetic field, lattices of stable magnetic whirls emerge \cite{seki2012formation,seki2012observation,nagaosa2013topological,langner2014coupledsublattices,zhang2016imagingdomains,zhang2016resonantxray,zhang2016multidomainSKX}. Due to the nanometer confinement and sensitivity to electromagnetic control, these magnetic structures have potential for new spintronic devices.

In particular, the spin direction of a magnetic skyrmion wraps once around the unit sphere. This implies that the spin configuration cannot be continuously deformed to another magnetic state i.e,  skyrmions are topologically protected particles \cite{nagaosa2013topological,muhlbauer2009skyrmion}. 
Therefore they can be created and destroyed only by singular magnetic configurations  \cite{milde2013unwinding}. Furthermore, for sufficiently small temperatures and magnetic fields, skyrmions can have practically an infinite lifetime \cite{hagemeister2015stability}. As a very smooth magnetic configuration, skyrmions couple only weakly to local defects \cite{muller2015capturing,iwasaki2013universal}. 
Instead, they couple extremely well to external forces. 
In metals and heterostructures, skyrmions can be manipulated by small electric or thermal currents \cite{iwasaki2013universal,jonietz2010spin,yu2012skyrmion,lin2014ac} while in magnetoelectric insulators they can be manipulated dissipationlessly with electric fields \cite{seki2012magnetoelectric,white2012electricfield,white2014electric,omrani2014exploration}, heat currents \cite{lin2014ac,koshibae2014creation}, and laser pulses \cite{ogawa2015ultrafast,Ghil2017}.
Furthermore, skyrmions are repelled from the edges of nanostructures and react very fast to external control \cite{sampaio2013nucleation,woo2016observation,zhang2015skyrmion,zhang2017motion}. 

From the application perspective, research has mainly focused on the manipulation of skyrmions in a ferromagnetic background. 
For example, it has been suggested to build memory devices \cite{fert2013skyrmions} based on skyrmions arranged in a nanowire where information is encoded in the distance between skyrmions. 
Alternative designs use nanostructures with lanes for skyrmions where the information is stored in the lane number \cite{muller2016magnetic}.

\begin{figure*}
	\includegraphics[width=0.24 \textwidth]{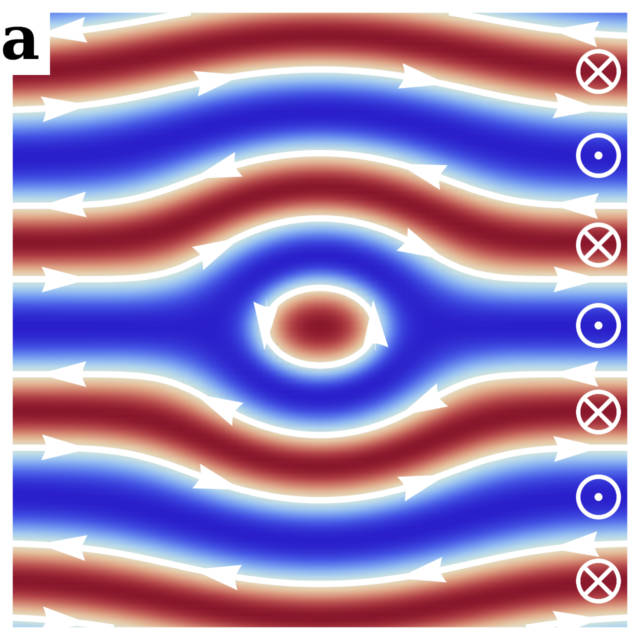}
	\includegraphics[width=0.24 \textwidth]{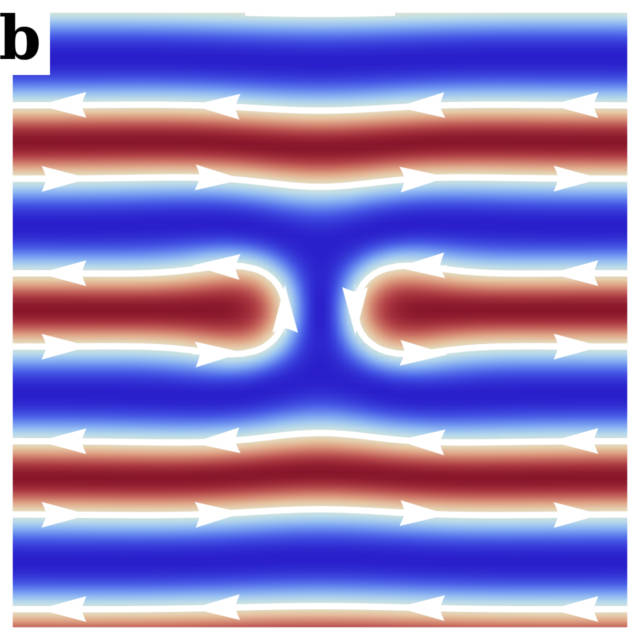}
	\includegraphics[width=0.24 \textwidth]{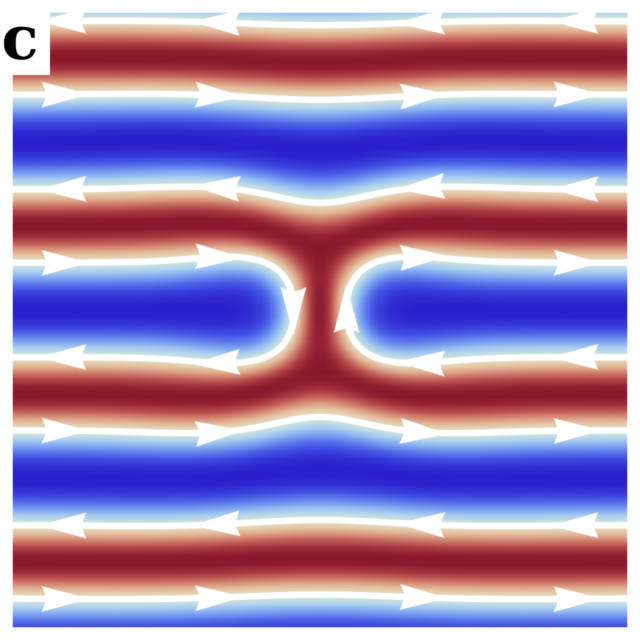}
	\includegraphics[width=0.24 \textwidth]{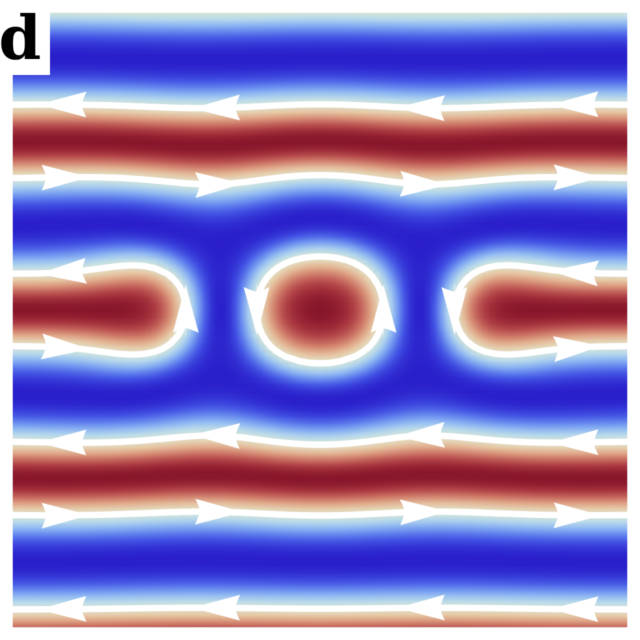} \\[1mm]
	\includegraphics[width=0.24 \textwidth]{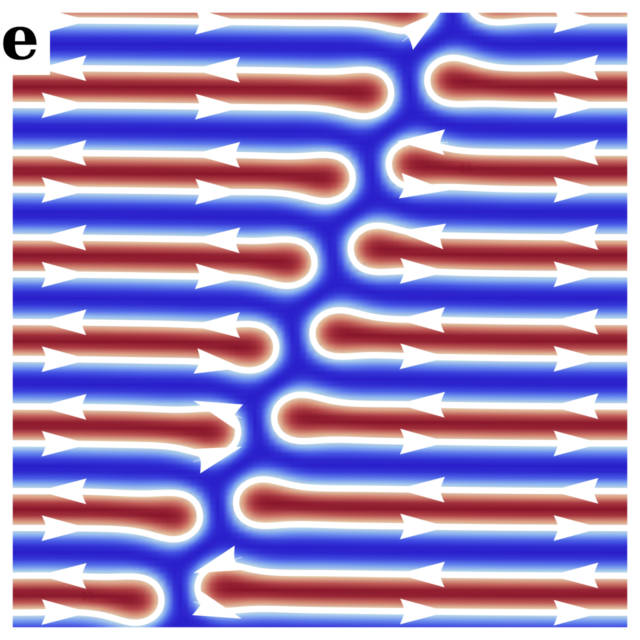} 
	\includegraphics[width=0.24 \textwidth]{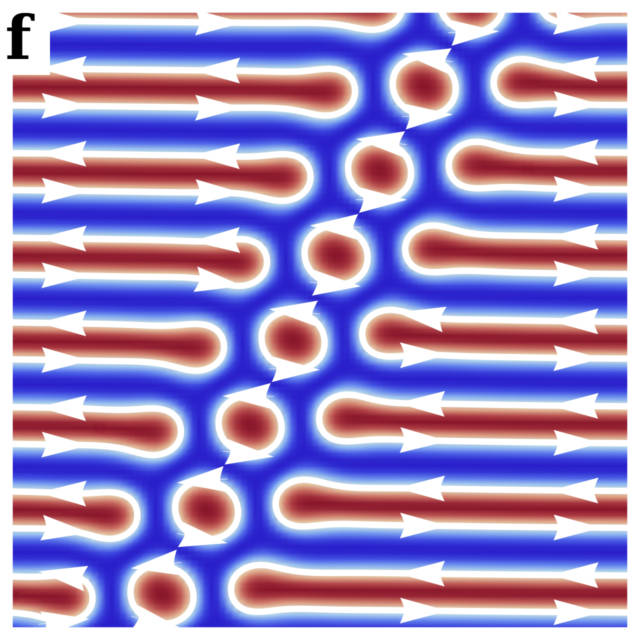} 
	\includegraphics[width=0.24 \textwidth]{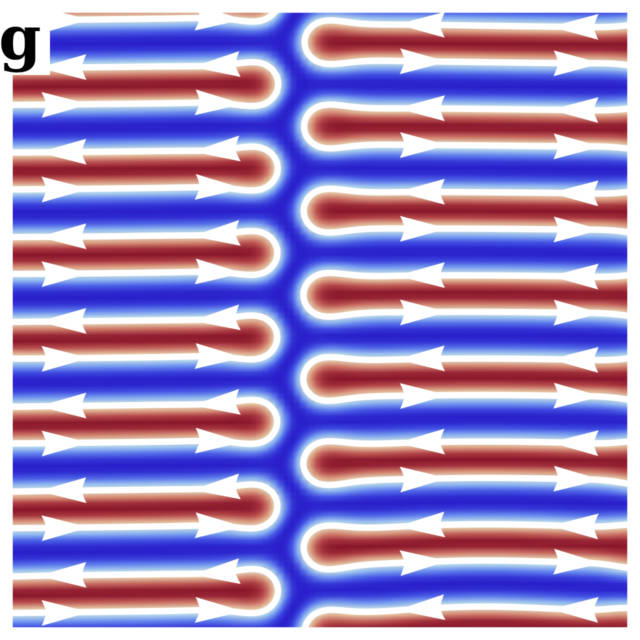} 
	\includegraphics[width=0.24 \textwidth]{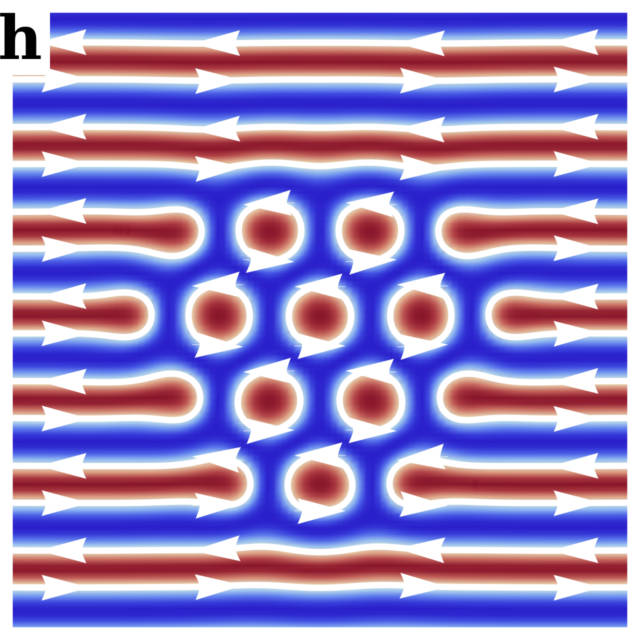} 
	
	\caption{ 
		Skyrmions in the helical phase ($B=0.2 B_0$). 
		The colorcode denotes the out-of-plane component of the magnetization as shown in (a).
		The in-plane component is sketched by white arrows.
		(a) and (b) are topologically equivalent single skyrmion states with winding number $-1$ and can be smoothly transformed one into the other. 
		The energy of the H-shaped skyrmion (b) is, however, much lower compared to the ``interstitial'' skyrmion of panel (a).
		In (c) an antiskyrmion with opposite winding number is shown. 
		Due to attractive interactions, two skyrmions form a dimer bound state with winding number $-2$, shown in panel (d). 
		(e-h) are multiskyrmion bound states, see text.
	\label{fig1}}
\end{figure*}
An interesting alternative is to consider skyrmions in a helical background. This problem, first studied by Ezawa \cite{ezawa2011compact},
has, up to now, received little attention. Such states naturally occur when the phase transition from 
the skyrmion crystal phase to the helical phase is investigated \cite{milde2013unwinding,schutte2014dynamics}. We argue that four major properties distinguish skyrmions in a helical background from their cousins in a ferromagnetic environment:
(i) Skyrmions naturally move along the tracks defined by the helical order. 
Currents perpendicular to these naturally formed tracks are expected to drive skyrmions to very high speeds especially in materials with low Gilbert damping \cite{iwasaki2013universal,iwasaki2013current,sampaio2013nucleation}.
(ii) In the helical phase, skyrmions are metastable even for vanishing magnetic field. In contrast, skyrmions in the ferromagnetic state require either a sufficiently strong external magnetic field or sufficiently strong easy-axis anisotropy to guarantee local stablility \cite{ezawa2011compact,muller2016edge,kezsmarki2015neel}.
(iii) While in the ferromagnetic state only one type of skyrmion is preferred by the background magnetization \cite{koshibae2016theory}, in the helical phase skyrmions and antiskyrmions can coexist moving on different tracks in the helical background.
(iv) While the skyrmion-skyrmion interaction is typically repulsive in the ferromagnetic case \cite{bogdanov1995interaction,zhang2015skyrmion}, there are strong attractive components in the helical phase which naturally lead to the formation of skyrmion dimers and larger skyrmion clusters.
Recently, clusters of skyrmions were also predicted \cite{leonov2016threedimensional}  in a conical background. 
Leonov {\it et al.} ~\cite{leonov2016chiral} studied experimentally the hysteretic behavior in this regime and the formation of multidomain patterns. After submission of this manuscript, a study by Loudon {\it et al. } \cite{loudon2017cluster} reported the observation of skyrmion clusters within the conical phase. Other works suggest attractive skyrmions in a fully polarized background only if an additional frustrated exchange \cite{Rozsa2016attractiveskyrmions} is taken into account. Also clusters in a homogeneous background have been measured under strong geometrical confinement \cite{zhao2016direct}.
In a helical background, however, the attractive interaction intrinsically arises from the helical modulations and thus no additional extensions of the minimal model are required.

In the following, we will first describe the theory of skyrmions in the helical phase, their interactions, and the formation of clusters of skyrmions. 
Energies and magnetic fields will be measured in units of $E_0$ and $B_0$, respectively. These are set by the exchange coupling and the critical field stabilizing the ferromagnet, see the Supplemental Material \cite{supplement}.
To observe these clusters experimentally, we acquired real space and real time movies of the helical-skyrmion phase transition in a thin film of Cu$_2$OSeO$_3$ using Lorentz transmission electron microscopy (LTEM) \cite{supplement}. The movies reveal the nucleation of skyrmions in the helical background and their tendency to arrange in clusters of different configurations.

{\em Single skyrmions in a helical phase:}
To investigate the energetics of skyrmions in a helical background, we performed micromagnetic simulations, see Supplemental Material \cite{supplement}. 
Figure \ref{fig1} depicts the different skyrmion configurations: top panels represent single skyrmions and the characteristic skyrmion dimer. Bottom panels show the different multiskyrmion cluster configurations. 
While the ``interstitial'' skyrmion in panel (a) resembles the well-known skyrmion in a ferromagnetic background, the H-shaped skyrmion of panel (b) can be viewed as a bound state of two half-skyrmions (merons) defined by the ends of a helical strip, also refered to as a {\it meron pair} in Ref.~\citenum{ezawa2011compact}. An antiskyrmion (panel (c)) with opposite winding number can be obtained by time reversal. Panel (d) represents the formation of a two-skyrmion dimer bound state due to the presence of attractive interactions.
Plots of the winding number densities show that the skyrmions are indeed well localized, see Supplemental Material \cite{supplement}.

\begin{figure*}	
	\includegraphics[width=0.47 \textwidth]{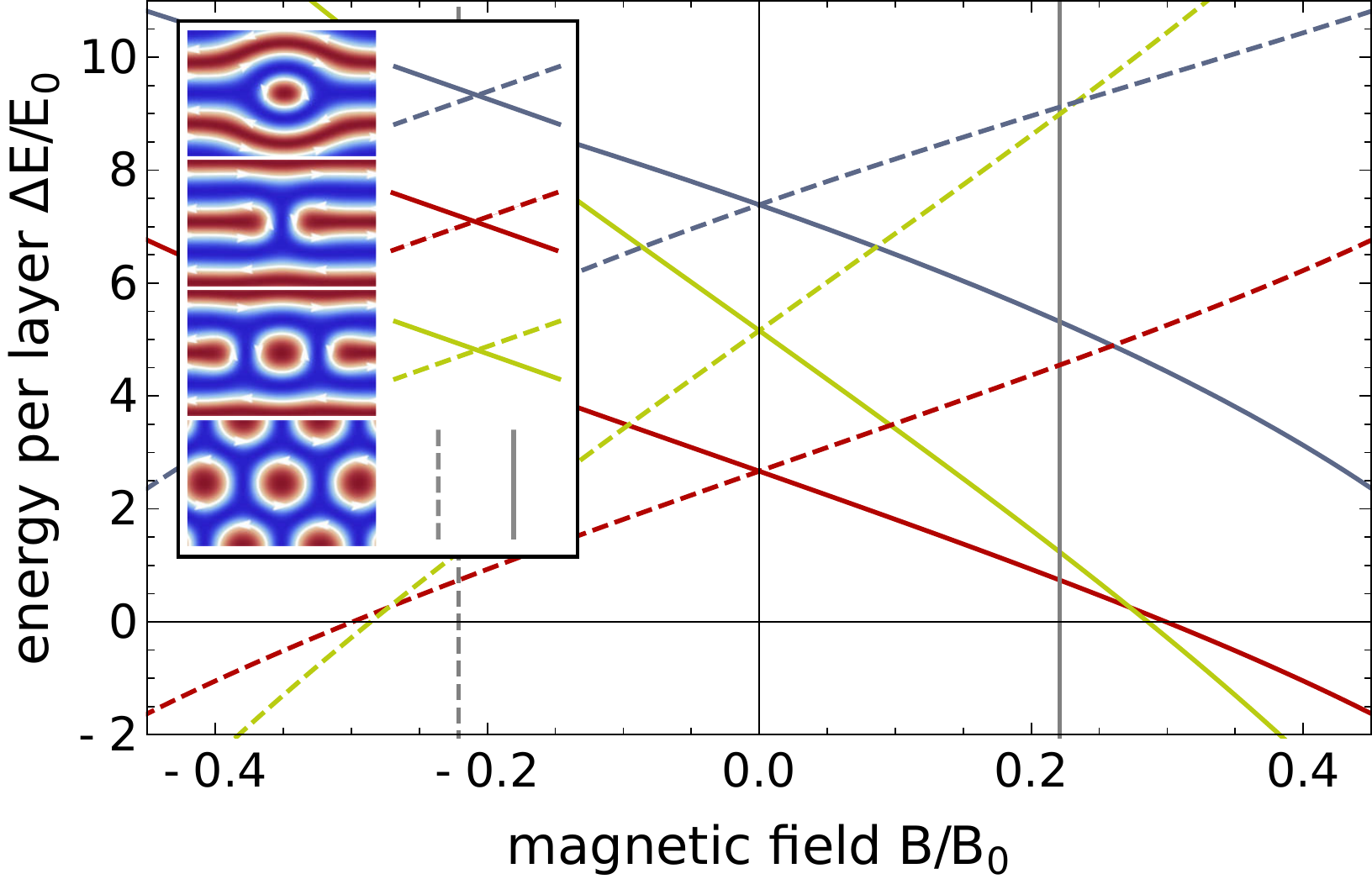}
	\hspace{2mm}
	\includegraphics[width=0.47 \textwidth]{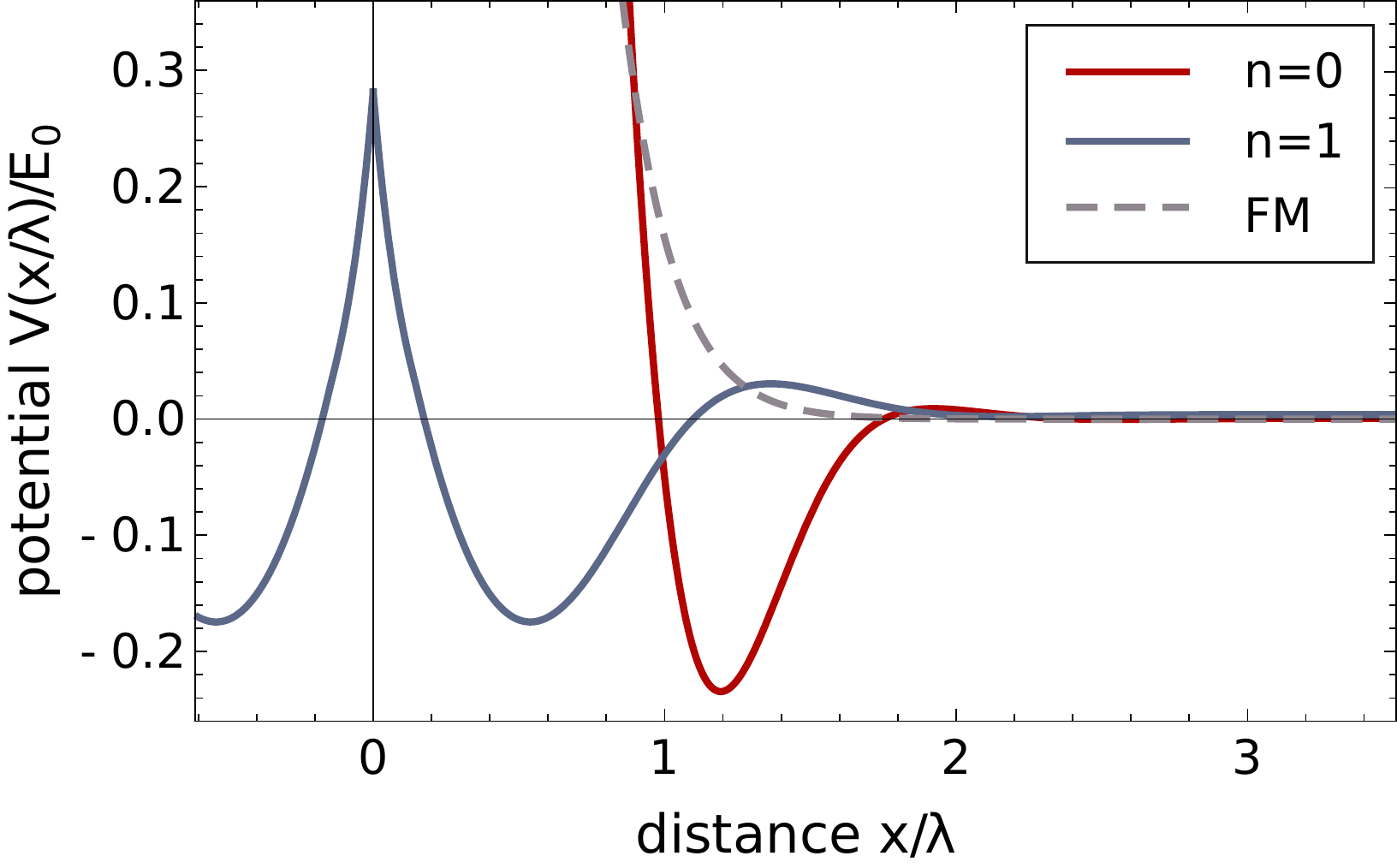}
	
	\caption{Energetics of the skyrmion configurations and effective interaction potentials.
		Left: The H-shaped skyrmion of Fig.~\ref{fig1}(b) has always a much lower energy (red) than the ``interstitial'' skyrmion of Fig.~\ref{fig1}(a) (blue). 
		The energy (light green) of the dimer (2-skyrmion bound state), Fig.~\ref{fig1}(d), is slightly lower than the energy of two separate skyrmions due to attractive interactions.
		The thermodynamic phase transition to the skyrmion lattice is indicated with vertical lines (gray).
		Dashed lines show the energies of the corresponding antiskyrmion states.
		Right: Interaction potentials at $B=0.2 B_0$ of two skyrmions on the same track ($n=0$) and on neighboring tracks ($n=1$) as a function of their distance ($\lambda$ is the wavelength of the helix at $B=0$). 
		For comparison, also the purely repulsive potential of two skyrmions in a polarized background at $B=1 B_0$ is shown ($FM$).
		Skyrmion-dimer and dimer-dimer interactions can to a good approximation be obtained by adding monomer-monomer potentials, see Supplemental Material \cite{supplement}. 
	\label{fig2}}
\end{figure*}
Figure \ref{fig2}(a) shows the energetics of four configurations of skyrmions (solid lines) and the corresponding antiskyrmions (dashed lines). 
Blue and red lines represent the energetics of the “interstitial” and the H-shaped configuration, respectively, and the green lines describe the dimer state.
The H-shaped state requires much less distortion of the helical background lattice and has a significantly lower energy than the interstitial state.
However, both states have the same winding number \cite{ezawa2011compact,supplement} and can be smoothly deformed into each other. 
For $B=0$, skyrmions and antiskyrmions are degenerate but since each skyrmion configuration carries a finite magnetization, they split linearly in the magnetic field. 
Above a critical field $B^c_s$ the energy of a single skyrmion becomes negative with respect to the pure helical background implying that above this field the system energetically favors the proliferation of skyrmions.
The exact critical field for the phase transition to the skyrmion lattice is, however, slightly lower due to attractive interactions between skyrmions (explained below).

By translational symmetry it costs no energy for a single skyrmion to move parallel to the track defined by the helical background. 
In contrast, a huge energy barrier of several exchange coupling constants (per layer of the material) prohibits the motion to the parallel lane as can be estimated from the large energy value of the interstitial skyrmion configuration.
This confinement has also important consequences for the velocity of the skyrmion when it is driven by a spin-current perpendicular to the confining walls where we assume that the helix is pinned by disorder. 
In a ferromagnet, the skyrmion would flow in the direction of the spin current with a velocity of the order of $v_s$, the velocity characterizing the spin current \cite{thiele1973steady}. 
Instead, the helical background acts as a confining potential \cite{iwasaki2013universal,iwasaki2013current,sampaio2013nucleation,muller2016magnetic,muller2015capturing} and accordingly one finds a velocity of the order of $v_s/\alpha$, if $v_s$ has a component perpendicular to the track. 
Here $\alpha$ is the Gilbert damping constant, which can be smaller than $10^{-2}$ in insulating materials like Cu$_2$OSeO$_3$ \cite{onose2012observation,heinrich2011spin}. 
This effect is somehow similar to a sailing boat which can obtain velocities larger than the velocity of the wind: in this analogy the keel of the boat takes over the role of the helical background.

{\it Interactions and cluster formation:}
Skyrmion-skyrmion interactions determine the collective behavior of skyrmions and the nature of phase transitions. Attractive interactions, for example, induce the formation of bound states and clusters and render the phase transition into the skyrmion phase to be of first order.
To investigate collective skyrmion states in the helical background, we calculate \cite{supplement} the skyrmion-skyrmion interaction potential $V_i(R)$ for two skyrmions in the H-type configuration and the results are shown in Fig.~\ref{fig2}(b). 
Red and blue solid lines describe the potential for skyrmions on the same track and on neighboring tracks, respectively.  
In both cases, the interaction is characterized by a weak long-ranged repulsion and a much stronger short ranged attraction. 
This leads to the formation of a bound state of two skyrmions, the dimer state depicted in Fig.~\ref{fig1}(d). 
For comparison, we also show the interaction potential for two skyrmions in a polarized background \cite{bogdanov1995interaction,zhang2015skyrmion} (gray dashed line).
Note that this potential is purely repulsive, which is in sharp contrast to the situation in a helical background.
The attractive potential in the helical background is strongest when the two skyrmions are on the same track (n = 0).
Remarkably, the attractive potential on neighboring tracks, $-0.175 E_0$, is only about 25\% lower than on the same track, $-0.234 E_0$. 
The minimal energy is thereby  obtained for a relative displacement of the two skyrmions of about half the helical wavelength either to the right or to the left. 
The attractive skyrmion-skyrmion interaction implies that it is energetically favourable to form clusters of skyrmions. 
Their energy can be estimated with high accuracy by adding the interaction potential of neighboring skyrmions. 
For $B=0.2 B_0$, for example, we obtain from this simple estimate $-0.469, -0.584, -4.67 E_0$ for the binding energy of three skyrmions in the same track, three skyrmions distributed on two tracks, 
and a 12-skyrmion cluster shown in Fig.~\ref{fig1}(h). 
These values can be compared to
$-0.482, -0.581, -4.60 E_0$ obtained from direct minimization. 
We have checked that similar results hold for $B=0.4 B_0$.

\begin{figure*}
	\includegraphics[width=1.0\textwidth]{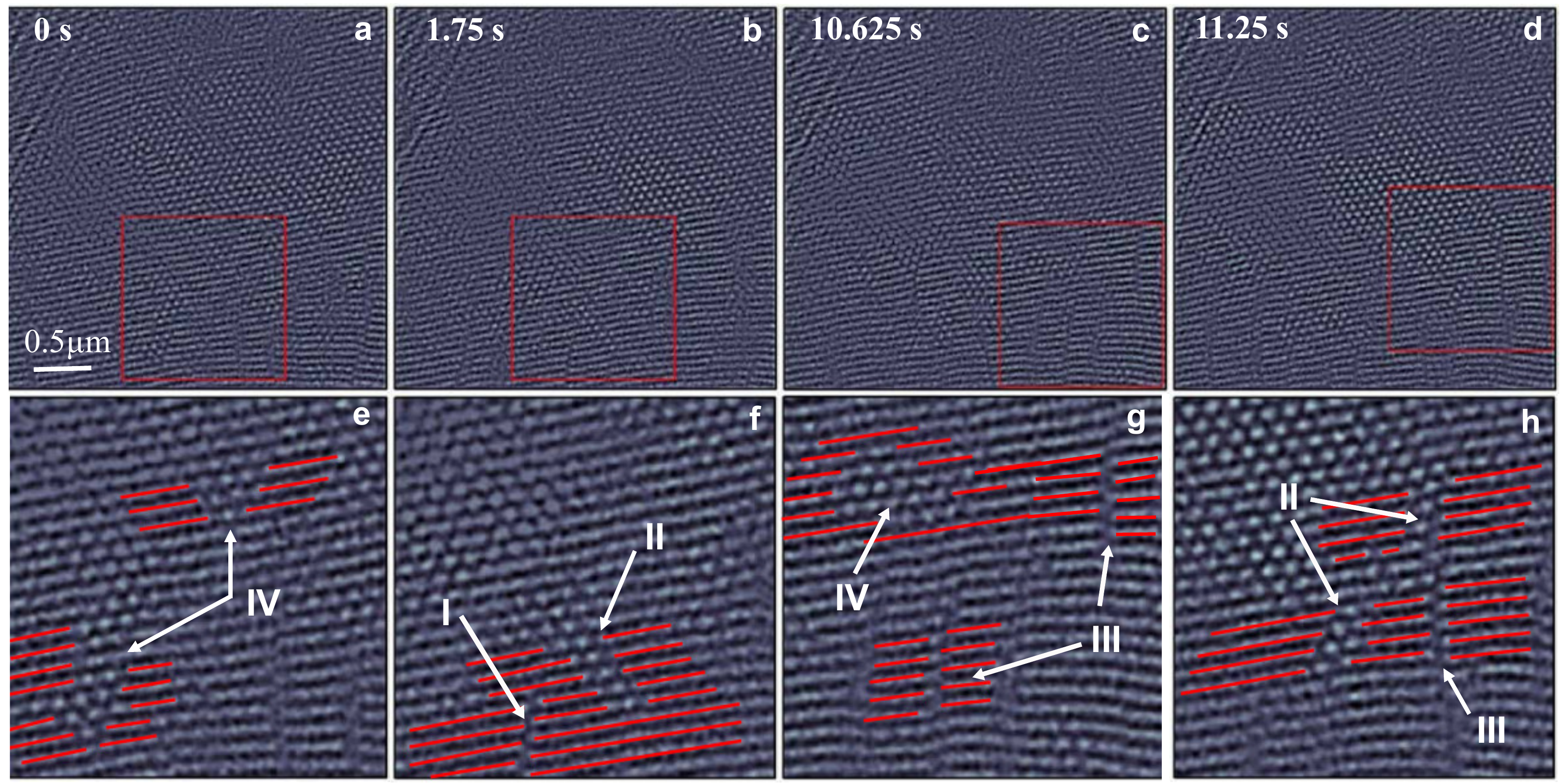}
	\caption{
		\label{fig3}
			Real-space cryo-Lorentz images reveal the coexistence of skyrmion and helical phases. Several types of skyrmion clusters, I - IV, can be identified corresponding to the simulated structures shown in Fig.~\ref{fig1}(e-h).} 
\end{figure*}

{\it Experiments:}
To experimentally detect the different skyrmion configurations predicted above, we have investigated a 110\;nm thin plate of Cu$_2$OSeO$_3$. The magnetic textures were imaged using a cryo-LTEM, see methods in the Supplemental Material. The applied magnetic field and temperature of the sample were held constant at 33\;mT and 13\;K, respectively, such that the system is close to the phase transition between the helical and the skyrmion phase. We recorded several movies and analyze here a selected movie of 100 frames with an integration time of 125\;ms per frame. 
We observe a coexistence of helical and skyrmion domains which fluctuate with time, see Supplemental Material \cite{supplement}.
Fig.~\ref{fig3}(a-d) represents a $3.3 \times 3.3$~$\mu$m$^2$ zoom of the total $5.7\times5.7$~$\mu$m$^2$ micrograph at four different time points. 
One can clearly identify large regions dominated either by skyrmion crystals or the helical phase. The interesting feature here is the appearance of smaller clusters of skyrmions inside the helical phases. For clarity, we show a zoom  of the real-space images marked by red squares in the lower panels (Fig.~\ref{fig3}(e-h)). To highlight the different configurations of skyrmions, some of the helical regions are marked in red.

We can identify several frequently occurring defect clusters that can be related to the theoretical states shown in Fig.\ref{fig1}(e)-(h). The configurations I forming a line of skyrmions correspond to the state shown in Fig.~\ref{fig1}(e). 
Also lines of dimers, compare Fig.~\ref{fig1}(f), are often found and labeled by II in Fig.~\ref{fig3}.
Averaging over several such configurations of I and II, we find that the angle between the defect orientation and the ordering vector of the helix is given by $23.5\pm 1.5^{\circ}$ for 
a line of skyrmions and $25.2\pm1^{\circ}$ for a line of dimers. 
This can be compared to  $24^\circ$ and $26^\circ$ obtained within our simulations.
We also find one further type of defect marked by III in Fig.~\ref{fig3} which cannot be interpreted as a bound state of skyrmions in a given helical background. 
Instead, in this ``zipper'' configuration the helical background on one side is shifted by half a lattice period. 
A simulation of such a structure is displayed in Fig.~\ref{fig1}(g). 
In our simulation the binding energy per skyrmion for this configuration is lowered by approximately $0.1 E_0$ compared to the skyrmion line of Fig.~\ref{fig1}(e) which implies that long defects of type I may transform into defects of type III.
Furthermore, we observe all types of irregularly shaped clusters denoted by IV in Fig.~\ref{fig3}. 
It is instructive to compare the formation of skyrmions in the helical and the ferromagnetic state. In the latter case, skyrmion-skyrmion interactions are repulsive \cite{bogdanov1995interaction,zhang2015skyrmion,muller2016magnetic} and therefore no formation of skyrmion clusters is expected. We have checked this by performing experiments at magnetic fields of 100\;mT to 150\;mT, where only skyrmions in a ferromagnetic background occur. As expected, no cluster formation was observed by us, see Supplemental Material \cite{supplement}.

We also observe regions where the image is blurred at the location of defects. We believe that this effect arises from the motion of defects. 
Assuming a Gilbert damping of the order of $10^{-2}$, we estimate the 1d diffusion constant of a cluster of $n$ skyrmions \cite{supplement} in the absence of defects to be of the order of $\frac{1}{n} 10^{-9}$\,m$^2$/s.  
Within the capturing time of a frame of our movie ($125 \textrm{ms}$) a freely moving cluster can therefore move distances of the order of $\frac{10}{\sqrt{n}}$ $\mu$m. 
This estimate is consistent with large fluctuations which we see from frame to frame captured within our movie. 
Sharply resolved images correspond to clusters which are temporarily bound to defects presumably arising either from lattice defects or height variations of our films.

{\em Conclusions:} While sharing the same topology, skyrmions embedded in a helical or a ferromagnetic environment have very different properties. Our studies highlight one aspect of this: attractive interactions in the helical background let skymions arrange in lines and clusters. These lines and clusters have characteristic shapes and orientations which we identified both in our micromagnetic simulations and electron microscopy experiments.
From the viewpoint of application, skyrmions in a helical background are potentially attractive since the helical stripes provide natural lanes along which skyrmions can be driven fast. However, one has to avoid the formation of zipper configuration which can, for example, be achieved in nanowires of suitable dimensions. The energy barrier for passing skyrmions between parallel lanes could be controlled for example with tailored light pulses or electric fields and will be explored in future experiments.

We thank the Deutsche Telekom Stiftung (J.M.), the  Bonn-Cologne Graduate School of Physics and Astronomy BCGS (J.M.) and the CRC 1238  (project C04) of the German Science Foundation (A. R., J.M.) for financial support.
J.R. thanks the Swiss National Science Foundation (SNSF) for funding through the Ambizione Fellowship PZ00P2\_168035.
Work at LUMES was supported by the National Center for Competence in Research Molecular Ultrafast Science and Technology (NCCR MUST), a research instrument of the SNSF. Work at LQM was supported by ERC project Controlled Quantum Effects and Spin Technology and SNSF (H.M.R.).

J. M., J. R. and P. H. contributed equally to this work.

\clearpage

\widetext
\begin{center}
\textbf{\large Supplemental Material for ``Magnetic skyrmions and skyrmion clusters in the helical phase of Cu$_2$OSeO$_3$''}

\vspace{4mm}
{Jan M\"uller,$^1$ Jayaraman Rajeswari,$^2$ Ping Huang,$^3$ Yoshie Murooka,$^2$ Henrik M. R\o nnow,$^3$ Fabrizio Carbone,$^2$ and Achim Rosch$\,^1$}

\vspace{1mm}
{\it\small $^1$Institute for Theoretical Physics, University of Cologne, D-50937 Cologne, Germany}

\vspace{-0.2mm}
{\it\small $^2$Laboratory for Ultrafast Microscopy and Electron Scattering (LUMES), Institute of Physics, EPFL, CH-1015 Lausanne, Switzerland}

\vspace{-0.2mm}
{\it\small $^3$Laboratory for Quantum Magnetism (LQM), Institute of Physics, EPFL, CH-1015 Lausanne, Switzerland}

\end{center}
\setcounter{equation}{0}
\setcounter{figure}{0}
\setcounter{table}{0}
\setcounter{page}{1}
\makeatletter
\renewcommand{\theequation}{S\arabic{equation}}
\renewcommand{\thefigure}{S\arabic{figure}}

\begin{center}
\small
\begin{minipage}[c]{0.8\linewidth}

\quad We present some details on the micromagnetic simulations, the calculation of interaction potentials and diffusion constants. Furthermore, we give experimental details and show that skyrmions close to the phase transition to the ferromagnetic state do not display the characteristic cluster formation observed in the helical phase.

\end{minipage}
\end{center}

\section{I. Micromagnetic Simulations}
We consider a two-dimensional plane with the magnetization $\mathbf{M}$ represented by a three-dimensional vector field.
The free energy functional, $F = \int d^2 \mathbf{r} \mathcal{F}$, includes exchange interaction $A$, Dzyaloshinskii-Moriya-interaction $D$ and an external magnetic field $B$:
\begin{equation}
\mathcal{F} = A (\nabla \hat{n})^2 + D \hat n \cdot \left( \nabla \times \hat n \right) - B M \hat n_z \text{,}
\label{eq:energy}
\end{equation}
where $\hat n = \mathbf{M}/M$ is the normalized magnetization.
Since the theory is continuous, we introduce dimensionless scales for the momentum $Q$, energy per layer $E_0$ and magnetic field $B_0$
\begin{equation} 
Q = \frac{D}{2A}, \quad  E_0 = 2 A, \quad B_0 = \frac{2 A Q^2}{M} \text{.}
\end{equation}
For numerical implementation, we discretize the system, typically using 22 spins per pitch $2 \pi/Q$ of the helix.
For the calculation of energy minima we prepare an initial magnetic texture and let it then relax by a fourth order Runge-Kutta integration of the equation
\begin{equation}
\partial_\text{t} \hat n = \mathbf{B_\text{eff}^\perp}
\label{eq:LLG}
\end{equation}
with $\mathbf{B}_\text{eff}^\perp$ the component of the effective magnetic field $\mathbf{B}_{\rm eff} = - \frac{1}{M} \delta F/\delta \hat n$ perpendicular to the magnetization.
This equation can be interpreted as the Landau-Lifshitz-Gilbert equation where the precession term is omitted.
For the calculation of interaction potentials we fixed a couple of spins (one to nine) in the center of the monomers during the minimization and hence ensure that the distance is preserved during the minimization process. We have checked that 
the results are approximately independent of the number of fixed spins.

\subsection{A. Localization of topological charge}

\begin{figure}[h]
	\includegraphics[width=0.24 \textwidth]{1a.jpg}
	\includegraphics[width=0.24 \textwidth]{1b.jpg}
	\includegraphics[width=0.24 \textwidth]{1c.jpg}
	\includegraphics[width=0.24 \textwidth]{1d.jpg} \\[1mm]
	\includegraphics[width=0.24 \textwidth]{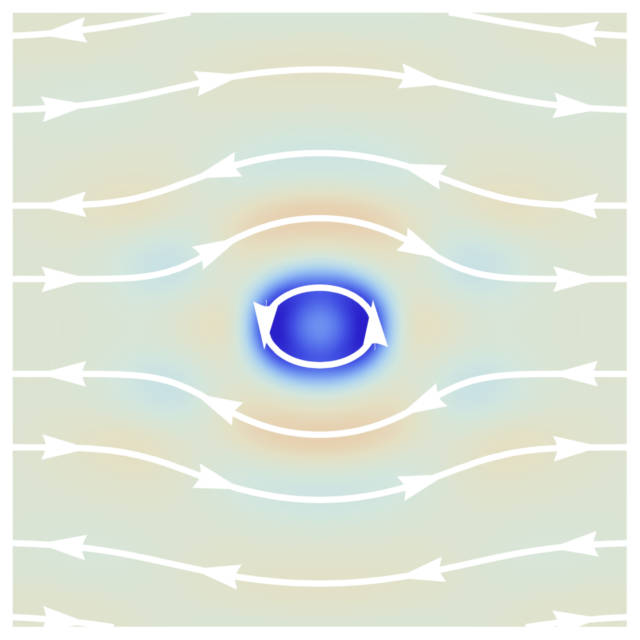} 
	\includegraphics[width=0.24 \textwidth]{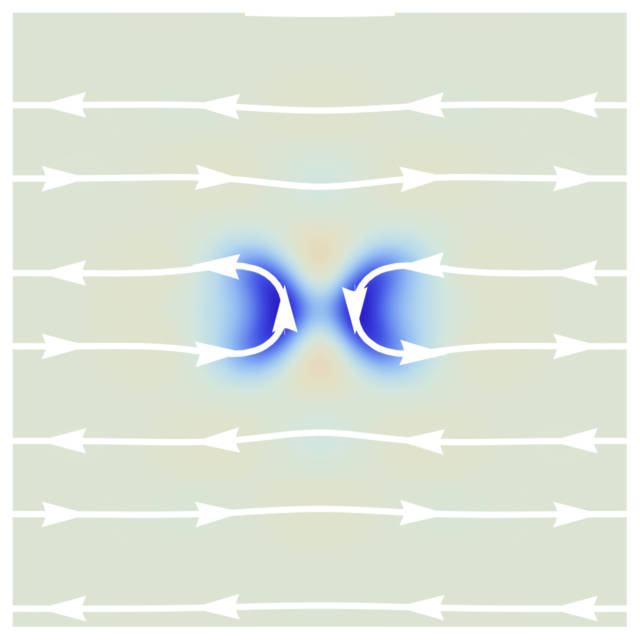} 
	\includegraphics[width=0.24 \textwidth]{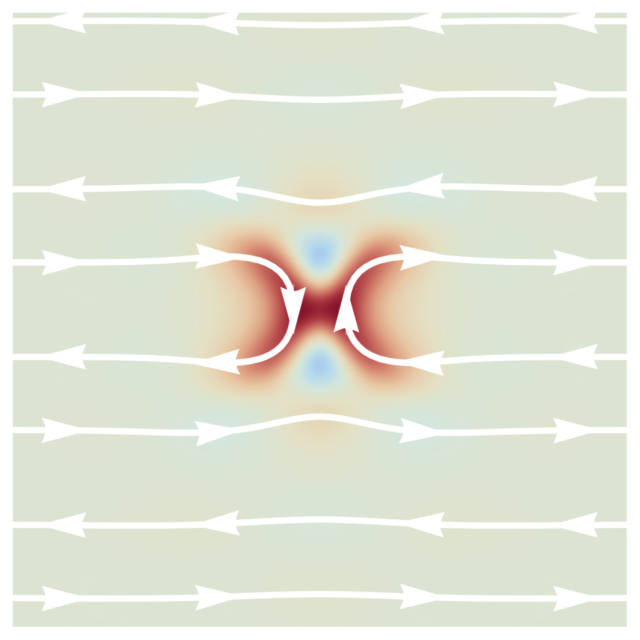} 
	\includegraphics[width=0.24 \textwidth]{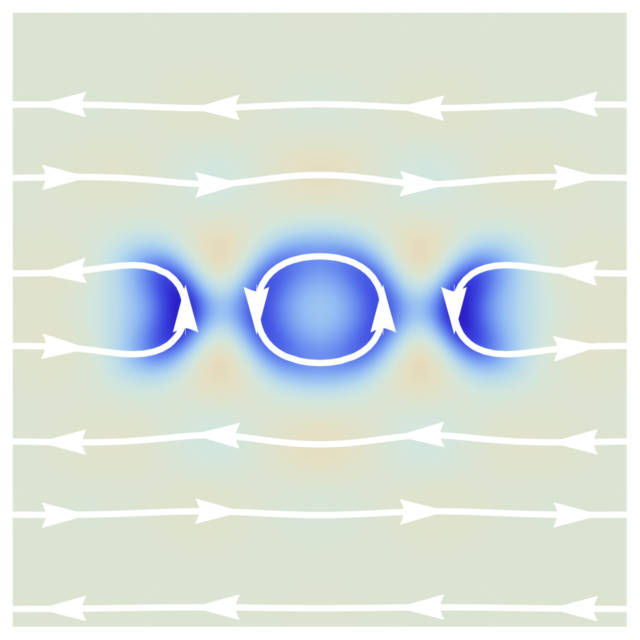} 
	
	\caption{ 
		Skyrmions in the helical phase ($B=0.2 B_0$) as also shown in the main text (upper row).
		The lower row of figures shows the corresponding winding number densities, rescaled to arbitrary units. The colorcode denotes the charge density as positive (red) or negative (blue) as compared to the uncharged helical state (pale green).
	\label{fig1}}
\end{figure}

We compute the topological charge density $w(\mathbf{r})$ from the relaxed monomer, anti-monomer and dimer configurations via the formula
\begin{equation}
w(\mathbf{r}) = \frac{1}{4\pi} \hat n (\mathbf{r}) \cdot \left( \partial_x n (\mathbf{r}) \times \partial_y n (\mathbf{r}) \right)
\label{eq:windingnumberdensity}
\end{equation}
and show the results in Fig.~\ref{fig1}.
What should be noticed here is that the charge distribution is indeed localized.
Furthermore, upon integration we find that the total charge $W = \int w(\mathbf{r}) \, \mathrm{d}^2r$ is $-1$ for the interstitial (a) and the H-shaped skyrmion (b), $+1$ for the antiskyrmion (c) and $-2$ for the skyrmion dimer (d).

\subsection{B. Dimer potentials and energetics of clusters}

\begin{figure}[h]
	\includegraphics[width=.6\linewidth]{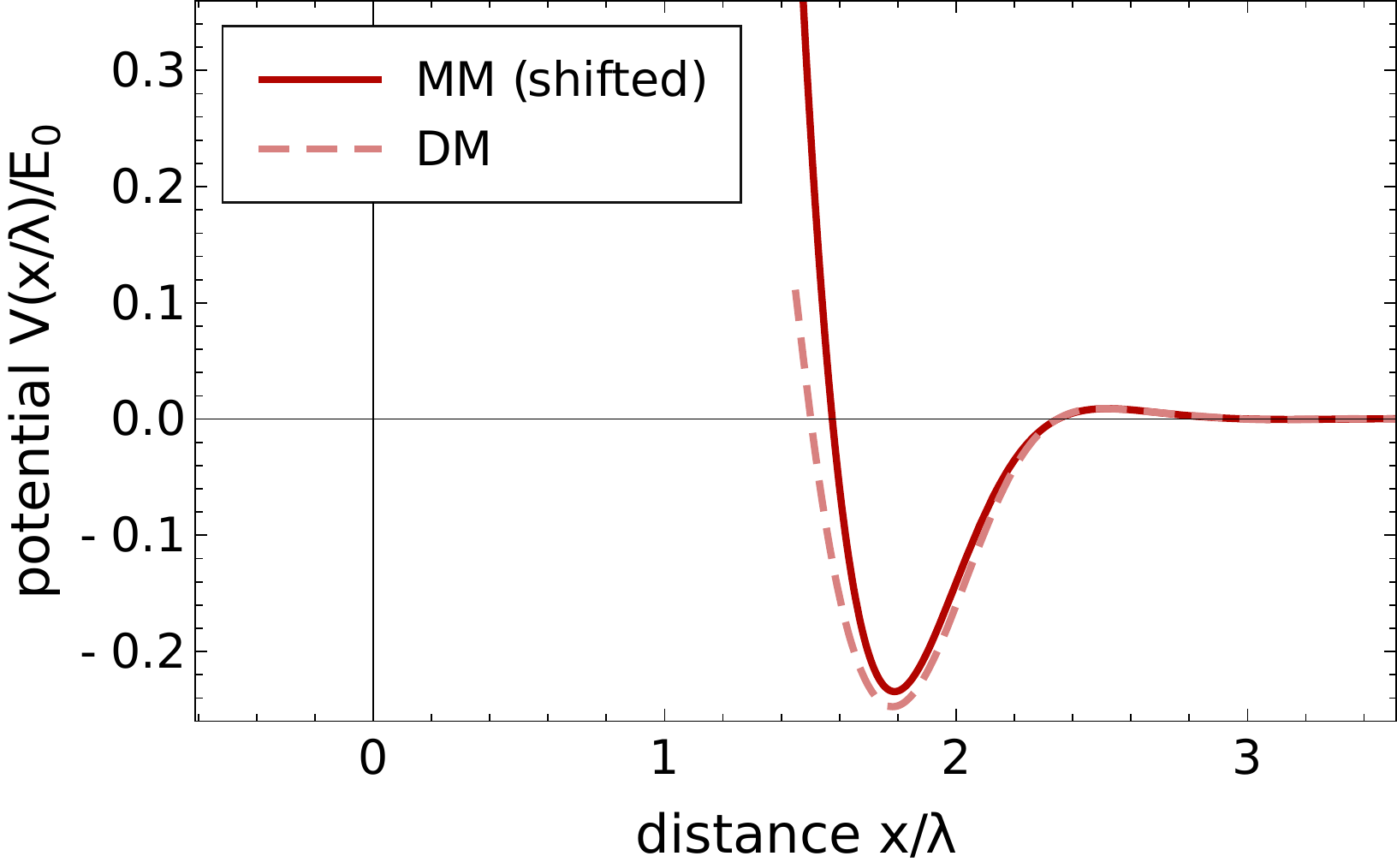}
	\caption{Interaction potential of a dimer with a monomer ({\it DM}, dashed) and a monomer with a monomer ({\it MM}, solid) on the same helical lane. The MM potential is shifted such that the position of one monomer matches the position of the central monomer in the DM potential. }
\label{fig4}
\end{figure} 

\begin{figure}[h]
	\includegraphics[width=.6\linewidth]{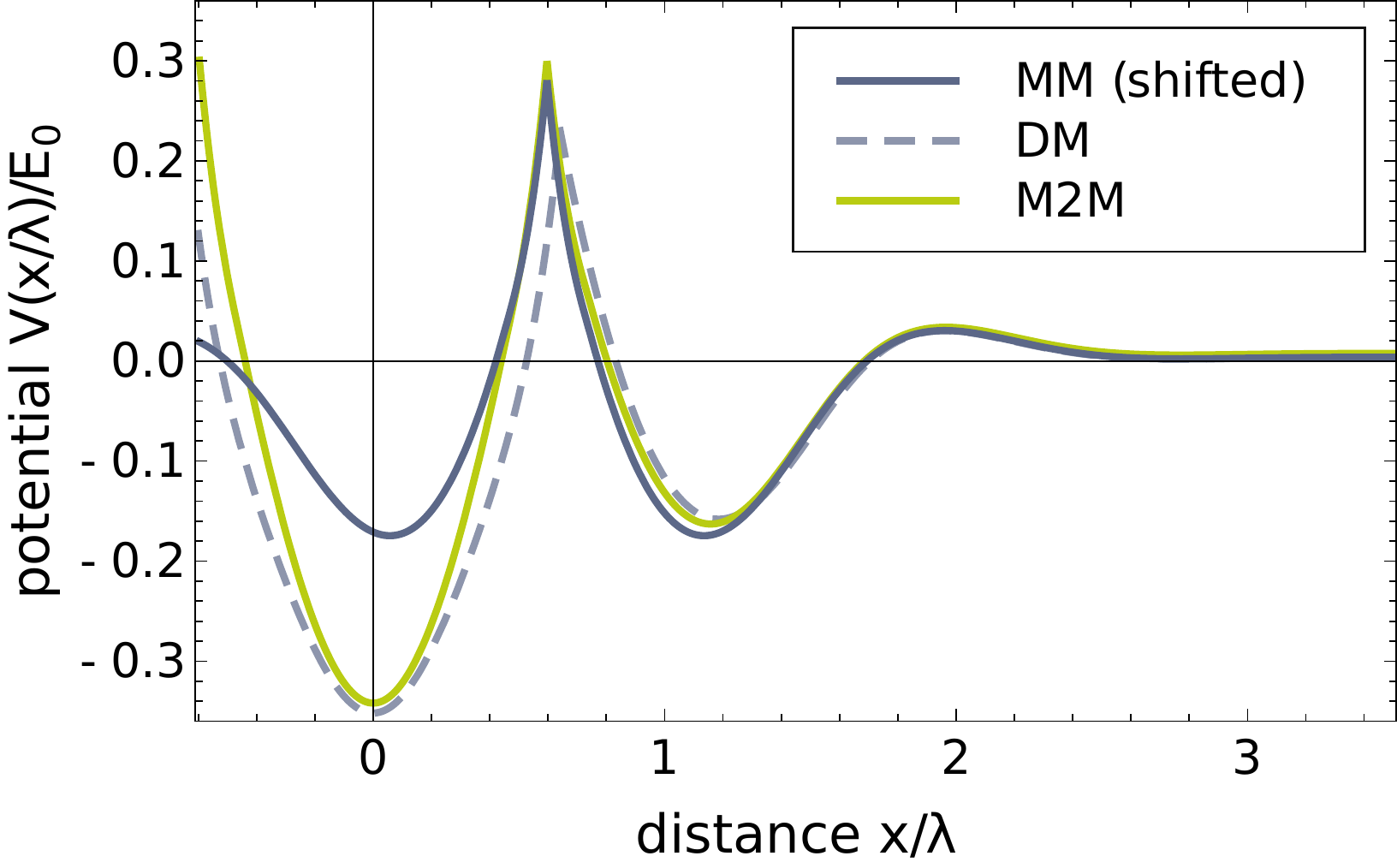}
	\caption{Interaction potential of a dimer with a monomer ({\it DM}, dashed) and a monomer with a monomer ({\it MM}, solid blue) on the same helical lane. The MM potential is shifted such that the position of one monomer matches the position of one monomer of the dimer in the DM potential. The green curve ({\it M2M}) shows a simple two-particle approximation of the DM potential. It is a superposition of twice the MM potential, shifted to the positions of both the monomers in a dimer.}
\label{fig5}
\end{figure} 

In the main text the monomer-monomer potentials are discussed.
Figures \ref{fig4} and \ref{fig5} show the interaction potential of a dimer (i.e. a bound pair of monomers) with a monomer on the same lane (Fig.~\ref{fig4}, red dashed) and on the neighboring lane (Fig.~\ref{fig5}, blue dashed).
We emphasize that these interactions can be reduced to the potentials of only monomer interaction by additionally plotting the interaction between monomers in the same setups (solid lines respectively).
These two-particle interaction potentials are shifted on the x-axis such that they match the positions of the single monomer and the nearest other monomer of the full three-particle potentials.
In the case of the dimer and the monomer being placed on neighboring lanes, Fig.~\ref{fig5}, the monomer can be located between the two constituents of the dimer and thus feels both of them. Here the full three-particle potential can be accurately described by a superposition of two two-particle interaction potentials accordingly shifted (solid green).

The energy of a cluster of skyrmions can then be estimated by simply adding the energies at the interaction minima. For example, the energy of the $12$-skyrmion cluster, Fig.~1(h) in the main text, is the sum of in total $8$ monomer pairs on the same lane and $16$ monomer pairs on neighboring lanes. This gives a total binding energy of $-4.67 E_0$ compared to $-4.60 E_0$ from the full relaxation.

\subsection{C. Diffusion}

The dynamics of the magnetization are governed by the Landau-Lifshitz-Gilbert equation (LLG).
In the presence of thermal fluctuations, these equations are made stochastic by adding a fluctuating term $\mathbf{b}^\text{th}(\mathbf{r},t)$ to the effective magnetic field $\mathbf{B}_\text{eff} \to \mathbf{B}_\text{eff} + \mathbf{b}^\text{th}(\mathbf{r},t)$.
This fluctuating field has to vanish on average
\begin{equation}
 \langle \mathbf{b}^\text{th}_i(\mathbf{r},t) \rangle = 0
\end{equation}
and is uncorrelated with a suitable normalization factor
\begin{equation}
 \langle \mathbf{b}^\text{th}_i(\mathbf{r},t) \mathbf{b}^\text{th}_j(\mathbf{r'},t') \rangle 
 = 2 \alpha \frac{k_B T}{\gamma M} \delta_{i j} \delta(r-r')\delta(t-t') \text{,}
\end{equation}
where $\alpha$ is the Gilbert damping, $\gamma$ the gyromagnetic ratio and $M$ the saturation magnetization.
Application of the Thiele ansatz, where the magnetization is assumed a rigid object, yields the thermal stochastic force $\mathbf{F}^\text{th}(t)$ on a skyrmion.
Its average properties follow directly from the LLG after the Thiele procedure
\begin{equation}
 \langle \mathbf{F}^\text{th}_i(t) \rangle = 0, 
 \quad 
 \langle \mathbf{F}^\text{th}_i(t) \mathbf{F}^\text{th}_j(t') \rangle = 2 \alpha \mathcal{D}_{i j} k_B T \delta(t-t'),
\end{equation}
where $\mathcal{D}_{i j} = s d \int \left(\frac{\mathrm{d}{\hat{\mathbf{M}}}}{\mathrm{d}r_i} \cdot \frac{\mathrm{d}{\hat{\mathbf{M}}}}{\mathrm{d}r_j}\right) \mathrm{d}r^2$ is the corresponding entry of the dissipation matrix, including the spin density $s$ and the thickness $d$ of the sample. $\hat{\mathbf{M}} = \mathbf{M}/M$ is the normalized magnetization.

Via the Thiele equation, the thermal force $\mathbf{F}^\text{th}(t)$ relates directly to the velocity $\dot{\mathbf{R}}$ of the skyrmion. The resulting mean squared displacement $\langle \Delta \mathbf{R}^2 \rangle_t$ for the free motion in two spatial dimensions follows to vanish linear in the damping $\alpha$,
\begin{equation}
 \langle \Delta \mathbf{R}^2 \rangle_t = 2 \frac{ 2 \alpha \mathcal{D}}{(\alpha \mathcal{D})^2 + \mathcal{G}^2}  k_B T t \text{,}
\end{equation}
which can be understood as the strong circular motion of the skyrmion due to a dominant Magnus force if the damping is low.
However, the motion of a skyrmion in the helical background is effectively confined to only one dimension.
In this case, a low damping results in a fast velocity since the mean squared displacement in the helical lane after a time $t$ follows as 
\begin{equation}
 \langle \Delta x^2 \rangle_t = 2 \frac{k_B T}{\alpha \mathcal{D}_{x x}} t = 2 D t
\end{equation}
where we obtained the diffusion constant $D = \frac{k_B T}{\alpha \mathcal{D}_{x x}}$.
Numerical evaluation of the dimensionless integral in the dissipation matrix coefficient yields $\mathcal{D}_{x x} \approx 65 s d$.
We used for the spin density $s= 1 \hbar / 89 \AA^3$, the sample thickness $d=150\mathrm{nm}$, the temperature $T=13K$ and estimated a Gilbert damping of $\alpha=0.01$.
Consequently, we obtain an estimate for the diffusion constant of the order $D= 10^{-9} m^2/s$.
Since the dissipation element $\mathcal{D}_{x x}$ is linear in the number of skyrmions $n$, the diffusion constant scales as $D \propto n^{-1}$.

\begin{figure*}
	\includegraphics[width=0.9\linewidth]{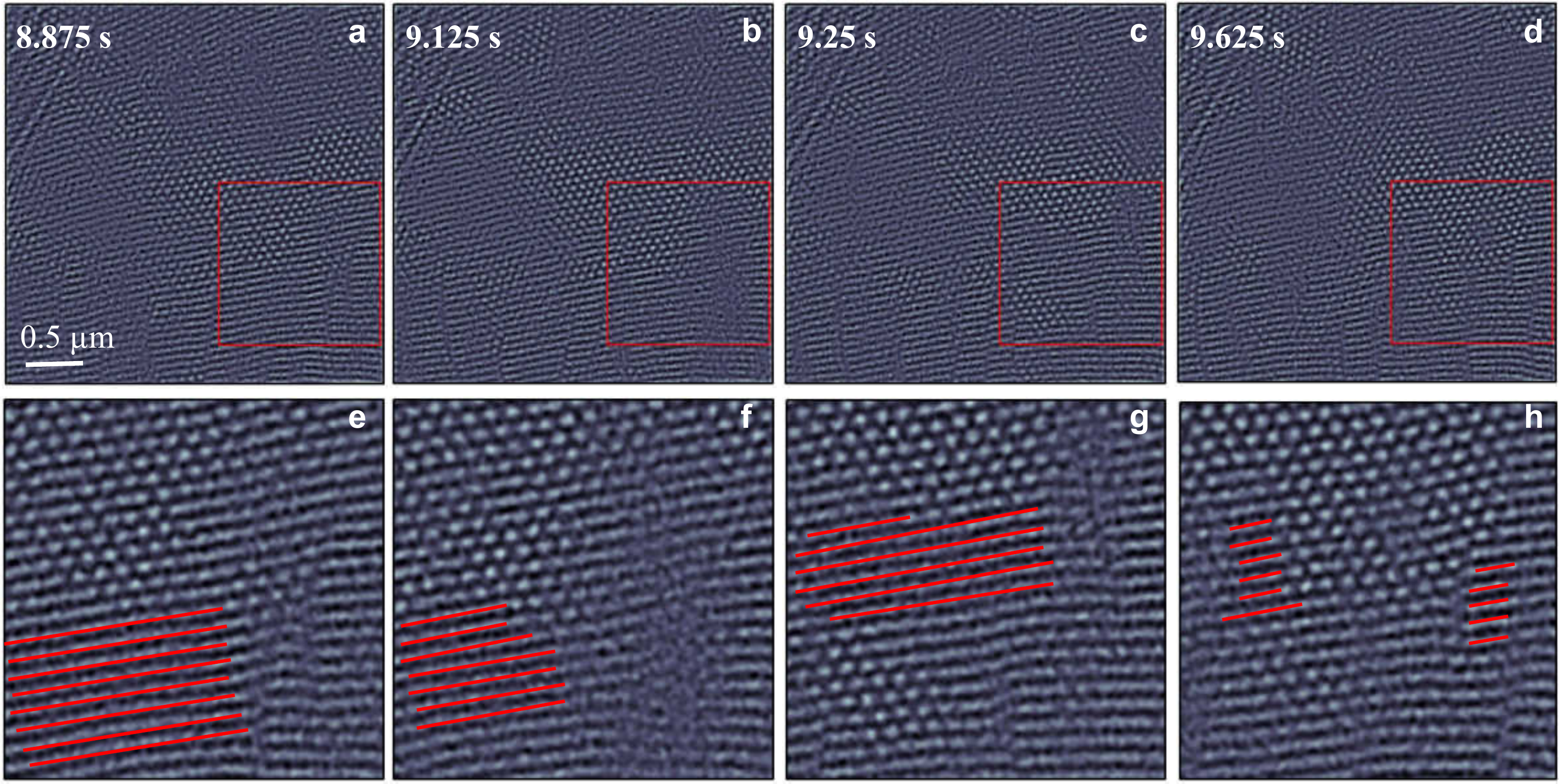}
	\caption{Four different frames of a moving depicting the energetic fluctuations between the skyrmion and helical phases.} 
\label{fig6}
\end{figure*}

\section{II. Experimental details}
A high-quality single crystal of Cu$_2$OSeO$_3$ was grown by the chemical vapor transport method. 
The single crystal was aligned and cut into a cube so that the three main directions correspond to $[1 \bar{1} 0]$, $[1 1 1]$ and $[\bar{1} \bar{1} 2]$, respectively. 
Then, choosing $(1 \bar{1} 0)$ as the main surface, the cube was cut into slices of $\approx 0.5$\;mm thickness. The sample was thinned to about 110\;nm by Focused Ion Beam (FIB) technique.

The magnetic structures of the sample were investigated using the FEI Titan Themis cryo-LTEM \cite{CIME} after zero-field cooling.  The microscope equipped with a field emission gun was operated at 300\;kV in the Fresnel mode. The magnetic field was applied normal to the sample surface along the $[1 \bar{1} 0]$ direction.

\section{III. Fluctuation of helical and skyrmion domains}

Figure\;\ref{fig6} presents four different frames from the movie indicating the coexistence as well as the fluctuation of helical and skyrmion domains. Magnified views of the highlighted red squares are shown in the bottom panels. For clarity, some of the helical regions are drawn in red to underline their movement as function of time.

\begin{figure*}
	\includegraphics[width=0.8\linewidth]{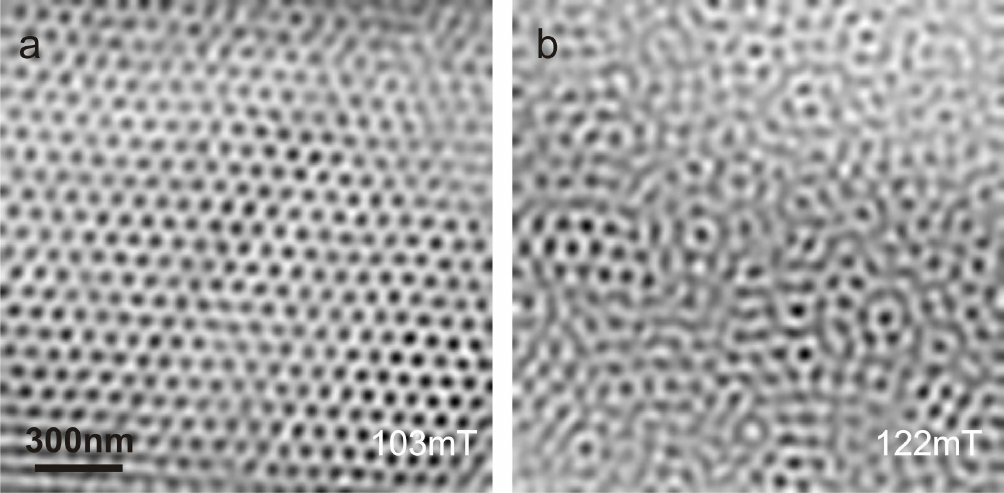}
	\caption{Lorentz TEM images depicting the skyrmion to ferrimagnet phase transition under increasing external magnetic fields.} 
\label{fig7}
\end{figure*}

\section{IV. Skyrmions in a ferromagnetic background}

Figure\;\ref{fig7} shows Lorentz micrographs upon increasing magnetic fields as the system passes from the skyrmion to the ferrimagnetic or conical phase. Since our method averages the signal over the sample thickness, we can not clearly distinguish between these two phases which occur at higher fields. As discussed in the main text, the interactions between skyrmions depend on the precise phase in which they are embedded. 
While skyrmions in the helical background form characteristic line defects, no such clusters can be found in Fig.~\ref{fig7}b, where skyrmions are embedded in a polarized background.

\section{V. Phase diagram}

\begin{figure*}
	\includegraphics[width=0.8\linewidth]{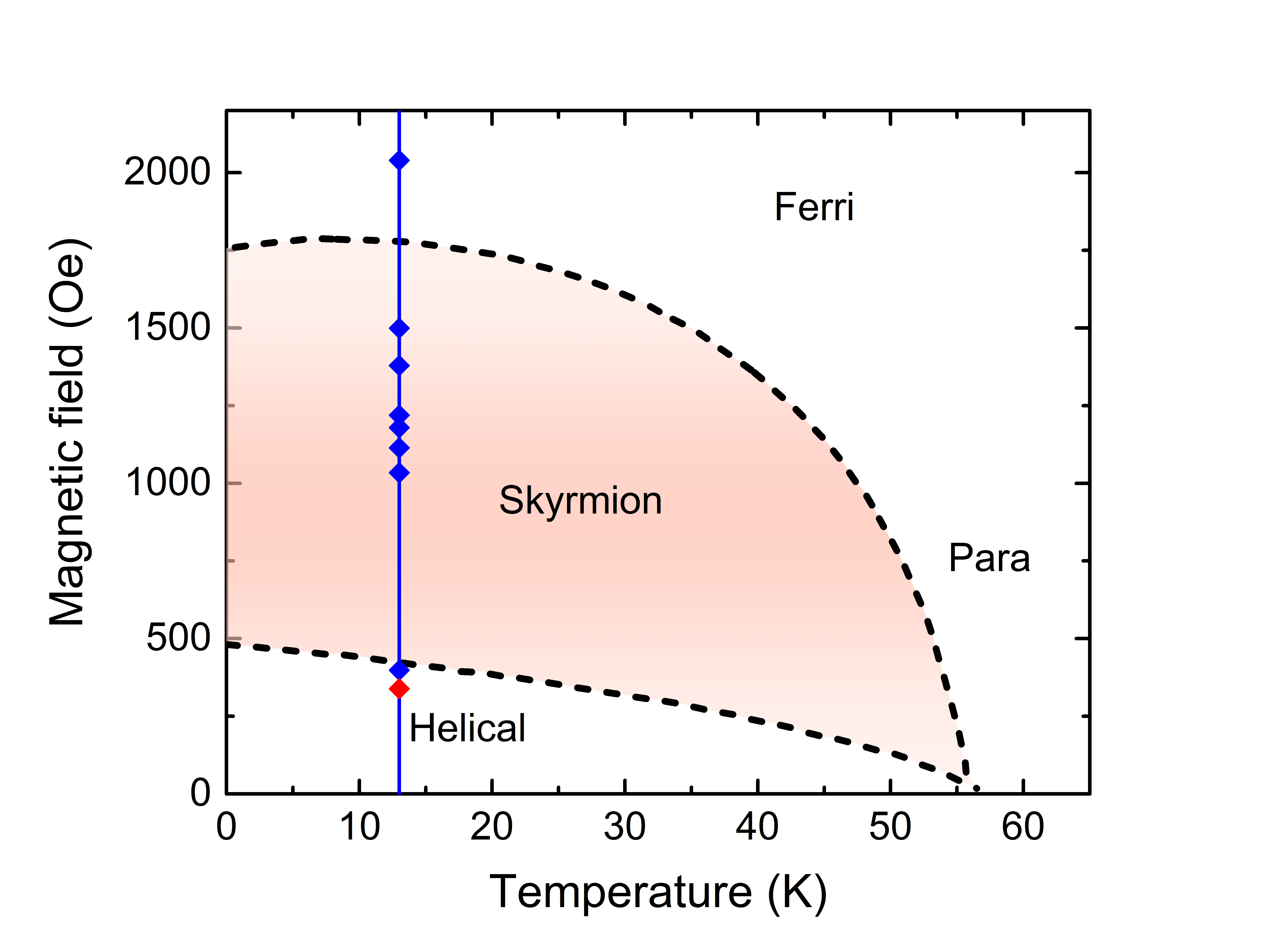}
	\caption{Schematic phase diagram for a $\approx 100\mathrm{nm}$ thin film of Cu$_2$OSeO$_3$. The parameters of our measurements are marked by diamond shaped dots. The red dot indicates the frames of the main text. The schematic phase diagram is a sketch based on the data points from Ref.~\citenum{seki2012observation}.} 
\label{fig8}
\end{figure*}

Figure\;\ref{fig8} shows a schematic phase diagram for a $\approx 100\mathrm{nm}$ thin film of Cu$_2$OSeO$_3$. The schematic phase diagram is a sketch based on the data points which we have taken from Ref.~\citenum{seki2012observation}. We have marked the positions in the phase diagram at which we record movies. The red diamond marks the point in phase space ($13$K, $33$mT) for which the movies in the main text were recorded. The blue diamonds mark other experimental parameters for which we have performed our measurements, among which are also the frames for the decay into the polarized phase, shown in Fig.~\ref{fig7}.

\end{document}
%